\documentclass[aps,prd,reprint,superscriptaddress]{revtex4-2}

\usepackage{subcaption}
\usepackage{orcidlink}

\begin{document}


\title{Scalable matched-filtering pipeline for gravitational-wave searches of compact binary mergers}


\author{Yun-Jing Huang \orcidlink{0000-0002-2952-8429}}
\email{yun-jing.huang@ligo.org}
\affiliation{Department of Physics, The Pennsylvania State University, University Park, PA 16802, USA}
\affiliation{Institute for Gravitation and the Cosmos, The Pennsylvania State University, University Park, PA 16802, USA}

\author{Chad Hanna}
\affiliation{Department of Physics, The Pennsylvania State University, University Park, PA 16802, USA}
\affiliation{Institute for Gravitation and the Cosmos, The Pennsylvania State University, University Park, PA 16802, USA}
\affiliation{Department of Astronomy and Astrophysics, The Pennsylvania State University, University Park, PA 16802, USA}
\affiliation{Institute for Computational and Data Sciences, The Pennsylvania State University, University Park, PA 16802, USA}


\author{Becca Ewing}
\affiliation{Department of Physics, The Pennsylvania State University, University Park, PA 16802, USA}
\affiliation{Institute for Gravitation and the Cosmos, The Pennsylvania State University, University Park, PA 16802, USA}

\author{Patrick Godwin \orcidlink{0000-0002-7489-4751}}
\affiliation{LIGO Laboratory, California Institute of Technology, MS 100-36, Pasadena, California 91125, USA}

\author{Joshua Gonsalves \orcidlink{0009-0000-2623-4356}}
\affiliation{Department of Physics, The Pennsylvania State University, University Park, PA 16802, USA}
\affiliation{Institute for Gravitation and the Cosmos, The Pennsylvania State University, University Park, PA 16802, USA}
\affiliation{Department of Computer Science and Engineering, The Pennsylvania State University, University Park, PA, 16802, USA}

\author{Ryan Magee \orcidlink{0000-0001-9769-531X}}
\affiliation{LIGO Laboratory, California Institute of Technology, Pasadena, CA 91125, USA}

\author{Cody Messick \orcidlink{0000-0002-8230-3309}}
\affiliation{Leonard E.\ Parker Center for Gravitation, Cosmology, and Astrophysics, University of Wisconsin-Milwaukee, Milwaukee, WI 53201, USA}

\author{Leo Tsukada  \orcidlink{0000-0003-0596-5648}}
\affiliation{Department of Physics and Astronomy, University of Nevada, Las Vegas, 4505 South Maryland Parkway, Las Vegas, NV 89154, USA}
\affiliation{Nevada Center for Astrophysics, University of Nevada, Las Vegas, NV 89154, USA}

\author{Zach Yarbrough \orcidlink{0000-0002-9825-1136}}
\affiliation{Department of Physics and Astronomy, Louisiana State University, Baton Rouge, LA 70803, USA}

\author{Prathamesh Joshi \orcidlink{0000-0002-4148-4932}}
\affiliation{Department of Physics, The Pennsylvania State University, University Park, PA 16802, USA}
\affiliation{Institute for Gravitation and the Cosmos, The Pennsylvania State University, University Park, PA 16802, USA}

\author{James Kennington \orcidlink{0000-0002-6899-3833}}
\affiliation{Department of Physics, The Pennsylvania State University, University Park, PA 16802, USA}
\affiliation{Institute for Gravitation and the Cosmos, The Pennsylvania State University, University Park, PA 16802, USA}

\author{Wanting Niu \orcidlink{0000-0003-1470-532X}}
\affiliation{Department of Physics, The Pennsylvania State University, University Park, PA 16802, USA}
\affiliation{Institute for Gravitation and the Cosmos, The Pennsylvania State University, University Park, PA 16802, USA}

\author{Jameson Rollins \orcidlink{0000-0002-9388-2799}}
\affiliation{LIGO Laboratory, California Institute of Technology, MS 100-36, Pasadena, California 91125, USA}

\author{Urja Shah \orcidlink{0000-0001-8249-7425}}
\affiliation{School of Physics, Georgia Institute of Technology, Atlanta, GW 30332, USA}


\date{\today}

\begin{abstract}
As gravitational-wave observations expand in scope and detection rate, the data analysis infrastructure must be modernized to accommodate rising computational demands and ensure sustainability. We present a scalable gravitational-wave search pipeline which modernizes the GstLAL pipeline by adapting the core filtering engine to the PyTorch framework, enabling flexible execution on both Central Processing Units (CPUs) and Graphics Processing Units (GPUs). Offline search results on the same 8.8 day stretch of public gravitational-wave data indicate that the GstLAL and the PyTorch adaptation demonstrate comparable search performance, even with float16 precision. Lastly, computational benchmarking results show that the GPU float16 configuration of the PyTorch adaptation executed on an A100 GPU can achieve a speedup factor of up to 169 times compared to GstLAL's performance on a single CPU core.

\end{abstract}


\maketitle

\section{\label{sec:intro}Introduction}

Since the first direct detection of gravitational waves (GWs) from a binary black hole merger, GW150914 \cite{gw150914}, the LIGO \cite{ligo}, Virgo \cite{virgo} and KAGRA \cite{kagra}\footnote{KAGRA joined midway through the third observing run.} collaborations have completed three observing runs, and reported over 90 candidates in the third Gravitational-Wave Transient Catalog (GWTC-3) \cite{gwtc-3}. These consistent detections not only affirm GW astronomy as an established field but also lead to groundbreaking discoveries, such as the first binary neutron star merger, GW170817 \cite{gw170817}. In light of upcoming ground-based and space-based detectors in multiple stages of development \cite{PhysRevD.91.082001,Punturo_2010,lisa}, we can expect a rapid increase in both data volume and detection rates, paving the way for new discoveries ahead.


As the scale of these observations expands, the computational demands on detection pipelines will increase drastically. 
GstLAL is a GW detection pipeline capable of processing data in low-latency \cite{messick2017, ewing2024, tsukada}. It has played a key role in the detection of GW150914 \cite{gw150914}, was the first pipeline to detect GW170817 in low-latency \cite{gw170817}, and has been consistently making detections throughout all the observing runs \cite{gwtc-1,gwtc-2,gwtc-2.1,gwtc-3}. GstLAL employs matched-filtering to search for compact binary coalescences (CBC) \cite{Cannon_2012}, a technique that is implemented in various forms in many search pipelines \cite{mbta, pycbc, spiir}. This technique is effective for identifying signals in noisy strain data, and requires filtering the data with a pre-generated bank of templates, typically containing O($10^6$) templates \cite{sakon}, which makes exploring high-dimensional parameter spaces computationally expensive.

Modern computational techniques, such as GPU acceleration and machine learning, have been explored for CBC searches \cite{spiir, aframe}. These approaches potentially provide scalability to handle increasing data size and allow for more extensive exploration of parameter spaces, leading to improved scientific insights. In this work, we adapt the core matched-filtering foundations of GstLAL to a modernized framework, PyTorch \cite{pytorch}, and demonstrate that it can maintain detection efficiency while providing the robustness needed to meet computational demands.

We chose the PyTorch framework for several reasons. First, the PyTorch library supports GPU execution, offering higher computational power and memory bandwidth, which allows for scaling the filtering pipeline and expanding both the template bank and search parameter space. Second, PyTorch enables seamless switching between CPU and GPU, supporting efficient use of both resources at computing centers while minimizing the need to maintain multiple code versions. Third, GPU support allows us to explore float16 precision, reducing computational costs and permitting further scaling of the template bank. Fourth, using Python, a high-level language, shortens development time, promoting faster progress and a broader developer and user community. Finally, PyTorch provides easy integration of machine learning techniques for future enhancements.

This paper is organized as follows: Section \ref{sec:pipeline} outlines the workflow of the GstLAL pipeline and introduces its adaptation to PyTorch. In Section \ref{sec:results}, we present search results from public GW data using the PyTorch adaptation and compare them to the original GstLAL results. Section \ref{sec:comp} evaluates and benchmarks the computational performance of the PyTorch adaptation against GstLAL. Finally, in Section \ref{sec:conclusion}, we discuss our conclusions and outline plans for future work.

\section{Pipeline description}\label{sec:pipeline}
The GstLAL search pipeline can be divided into two operational modes: ``online", which processes data in low-latency, and ``offline", which processes archival GW data. In this paper, we will perform analyses using the ``offline" mode of the GstLAL workflow. For a description of the ``online" mode of the GstLAL pipeline for O4, see \cite{ewing2024}. For a description of both online and offline search for O1 and O2, see \cite{messick2017}.


\begin{figure}[h]
    \centering
    \includegraphics[width=1\linewidth]{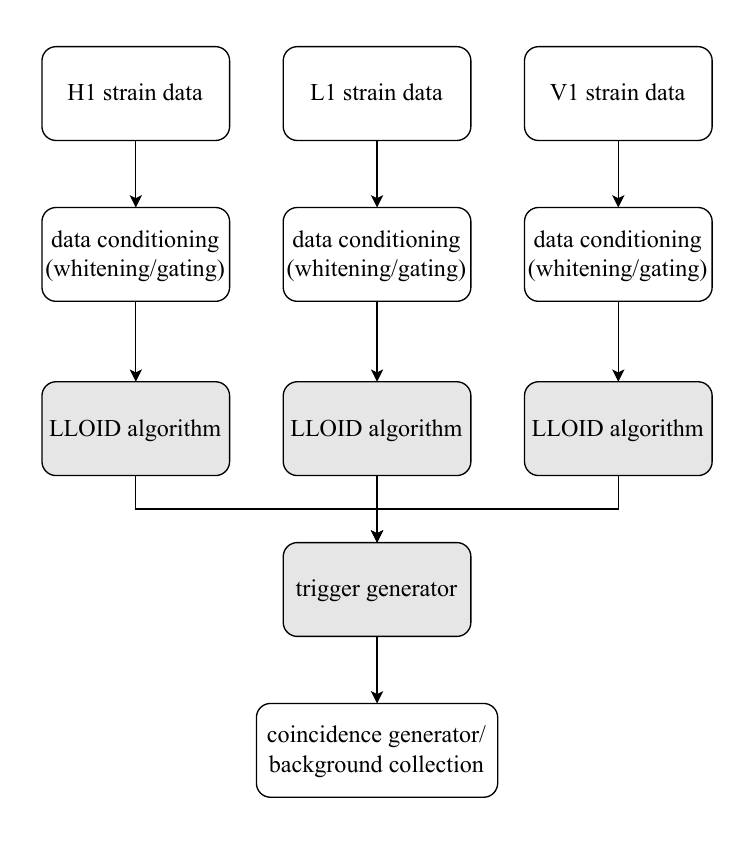}
    \caption{\label{fig:workflow}Flowchart of the inspiral program in the GstLAL offline workflow. The shaded elements are the processes adapted to the PyTorch framework, and can be executed either on the CPU or GPU, offering flexibility and portability. The workflow begins with strain data from multiple detectors being read into the pipeline. The data is then conditioned through whitening and gating. Afterward, the conditioned data is processed using the LLOID algorithm, GstLAL's implementation of matched-filtering \cite{Cannon_2012}. The LLOID algorithm generates SNR time series, and SNR peaks with values $\ge$ 4 are identified as triggers in the trigger generator. These triggers are then sent to the coincidence generator to form coincident triggers, while non-coincidence triggers during coincidence detector times are gathered as background data.}
\end{figure}

\subsection{GstLAL inspiral workflow}

GstLAL uses time-domain matched-filtering to search for GW candidates in detector strain data \cite{Cannon_2012,messick2017}. The process begins by correlating the data with a bank of templates, identifying potential candidates, and then assigning significance to each. The workflow can be broken down into several stages.

In the setup stage, the template bank for the matched-filtering process is generated. For the O4 run, the GstLAL template bank contains approximately $\sim2\times10^6$ CBC templates \cite{sakon}. These templates are divided into groups of $\sim1000$, and within each group, the Low-Latency Online Inspiral Detection (LLOID) algorithm is applied \cite{Cannon_2012}. The templates are split into different time slices, which then undergo singular value decomposition (SVD) \cite{svd}. Each group of decomposed templates is referred to as an ``SVD bank," which is used in the next filtering stage.

In the filtering stage, the strain data is processed by first estimating the Power Spectral Density (PSD). The running average of the PSD is used to whiten the strain data, after which large deviations in the whitened data are gated and removed, completing the data conditioning process \cite{messick2017}. The conditioned data is then filtered using the SVD bank from the setup stage, producing signal-to-noise ratio (SNR) time series for each template. This is achieved using the LLOID algorithm, which down-samples the conditioned data, cross-correlates it with orthogonal templates, performs matrix multiplication of the orthogonal SNR time series with SVD coefficients, and finally up-samples and sums the physical SNR segments \cite{Cannon_2012}. Once the LLOID algorithm produces the SNR time series, the inspiral program identifies peaks in the SNR time series with SNR $\ge$ 4, which are defined as ``triggers." The phase and the signal consistency metric $\xi^2$ are then calculated for these triggers \cite{cannon2015,tsukada}. These triggers are passed to the coincidence generator, which identifies coincidences between triggers detected with the same template and within the light travel time between detectors. Non-coincident triggers that occur during times when multiple detectors are observing are collected as background data.

In the injection stage, simulated CBC signals are generated based on a set of injection parameters, and these waveforms are added to the strain data at specific time intervals. The combined data is then filtered using the same algorithm applied in the filtering stage.

Finally, in the ranking stage, significance is assigned to the triggers identified during filtering. Triggers are ranked using a ranking statistic, which is then used to determine their false alarm rate (FAR). GstLAL uses the likelihood ratio as the ranking statistics, calculated against the background data collected during the filtering stage \cite{cannon2015,tsukada}. Triggers from the injection stage are also ranked against background data from the filtering stage. The fraction of injections that pass a given FAR threshold is then used evaluate the performance of the pipeline.

\subsection{PyTorch adaptation}\label{sec:pytorch}
The filtering stage of the GstLAL workflow, which filters whitened data with the template bank to produce SNR time series, is the primary computational bottleneck. In GstLAL, the LLOID algorithm is implemented as a set of Gstreamer \cite{gstreamer} elements in C, and runs exclusively on the CPU \cite{CANNON2021}. In this work, we re-implement the LLOID algorithm using PyTorch.

The LLOID algorithm performs several linear algebra operations including cross-correlation, matrix multiplication, resampling, and addition. In the PyTorch adaptation, we replace these operations with PyTorch library functions and integrate them into the workflow. Additionally, we introduce an optimization that enables filtering across multiple SVD banks in parallel. Since SVD time slices with the same sample rate share the same input and output dimensions during filtering, these templates are grouped and stacked together during initialization. We then leverage PyTorch's multi-batch and multi-channel operations to execute the filtering computations simultaneously across multiple SVD banks.

In addition to re-implementing the LLOID algorithm, the trigger generation stage is also adapted for PyTorch. This change is necessary because the LLOID algorithm generates SNR time series for each template in the bank, and transferring such a large dataset from the GPU to the CPU will create a substantial memory bandwidth bottleneck. By adapting the trigger generation to PyTorch, we can reduce the data transfer by creating a more refined dataset that is sent from the GPU to the CPU. The inspiral program in the GstLAL offline workflow and the PyTorch adaptation are shown in Figure \ref{fig:workflow}.

\section{O3 data search results}\label{sec:results}
In this section, we aim to demonstrate the validity of the PyTorch adaptation by comparing its offline GW search results with those of GstLAL. For the PyTorch adaptation, we show search results under three different configurations: (1) CPU with float32 precision, (2) GPU with float32 precision, and (3) GPU with float16 precision. First, we outline the dataset and search parameter space used for the analyses, followed by a comparison of the known GW events recovered by each search configuration. Finally, we evaluate the difference in injection recovery.

\subsection{Dataset}
We analyze an 8.8 day stretch of public strain data from the LIGO Hanford (H), LIGO Livingston (L), and Virgo (V) interferometers, obtained from the Gravitational Wave Open Science Center (GWOSC) \cite{gwosc}. The data cover the time range of May 12, 2019 19:36:42 UTC to May 21, 2019 14:45:08 UTC, which is during the third observing run (O3). 

\subsection{Search parameter space}
The template bank used in the analyses targets the BBH region and covers a parameter space with component masses between 3 and 200 $M_\odot$, chirp mass $M_{chirp}$ between 6 and 200 $M_\odot$, mass ratio between one and ten, and spin $z$-components between -0.35 and 0.35. The frequency range for the matched-filter integration is from 15 to 1024 Hz. In total, the bank consists of 9,693 templates, which are grouped into 10 SVD banks.

\subsection{Gravitational wave events}
There are six GW events previously reported in GWTC-3 within the time span of the dataset \cite{gwtc-3}. 
Using the small BBH template bank, GstLAL and all three different configurations of the PyTorch adaptation recovered the six known GW events consistently, which are summarized in Table \ref{table:events}.

The best match templates were the same among all four analyses for each of the six events.
The PyTorch adaptation recovers slightly lower network SNRs than those of GstLAL for the events GW190513\_205428, GW190514\_065416, GW190521, with a relative difference of 0.2\%, 0.2\%, and 0.01\%, respectively. This discrepancy can be attributed to the additional feature of sub-sample interpolation in the GstLAL pipeline during the SNR peak finding stage, which is not implemented in the PyTorch adaptation and is left for future development. The network SNRs for the other three events are consistent across all four analyses.
The FAR values recovered by the analyses are also consistent. The results for GPU float16 show that performing matched-filtering in half precision can produce consistent results with those of float32 precision, and can be effectively used for detection in a GW search pipeline.

\begin{figure*}
    \centering
    \begin{subfigure}{0.49\textwidth}
        \centering
        \includegraphics[width=\textwidth, keepaspectratio]{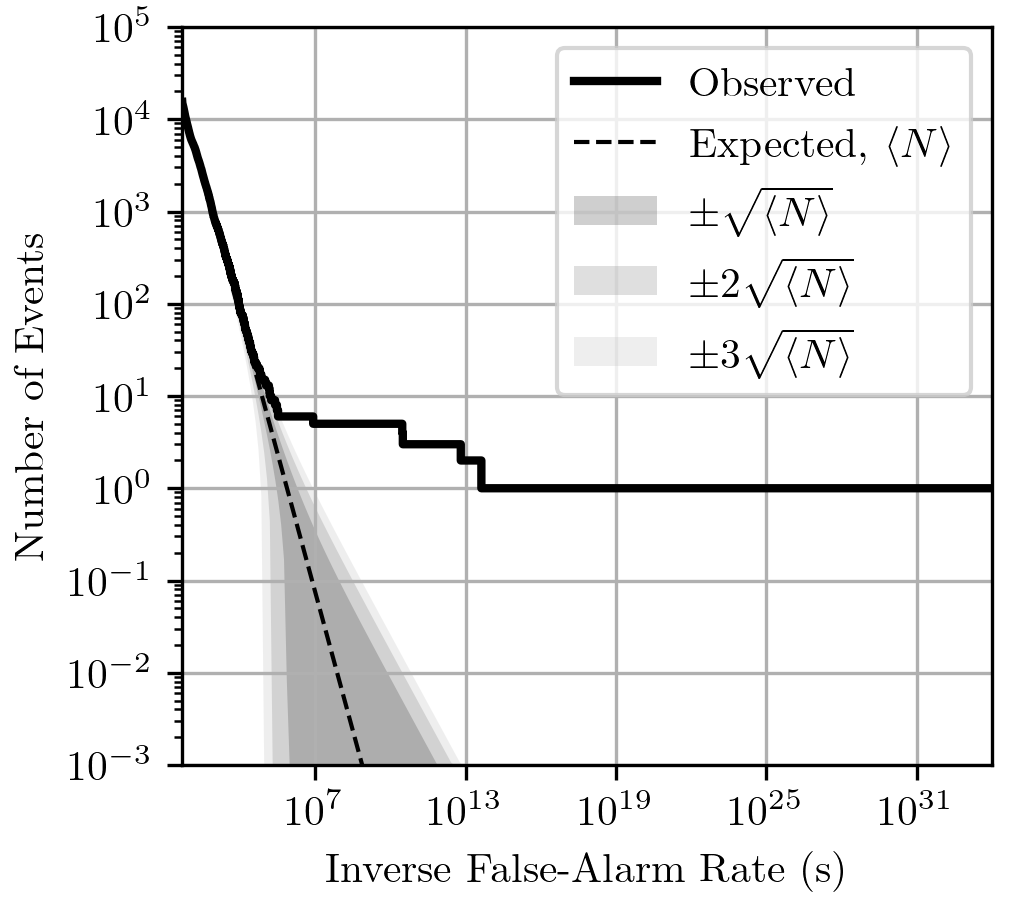}
        \caption{\label{fig:ifar:gstlal}GstLAL CPU float32}
    \end{subfigure}
    \hfill
    \begin{subfigure}{0.49\textwidth}
        \centering
        \includegraphics[width=\textwidth, keepaspectratio]{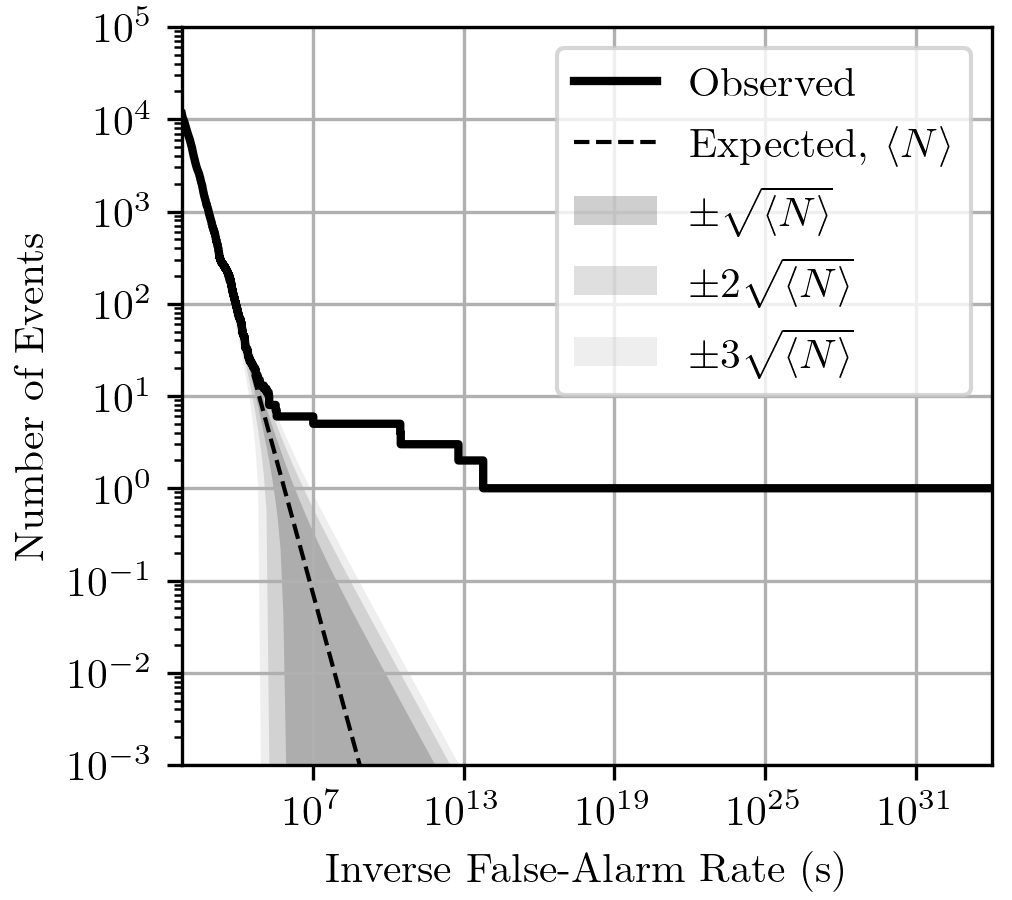}
        \caption{\label{fig:ifar:greg_cpu}PyTorch CPU float32}
    \end{subfigure} \\
    
    \begin{subfigure}{0.49\textwidth}
        \centering
        \includegraphics[width=\textwidth, keepaspectratio]{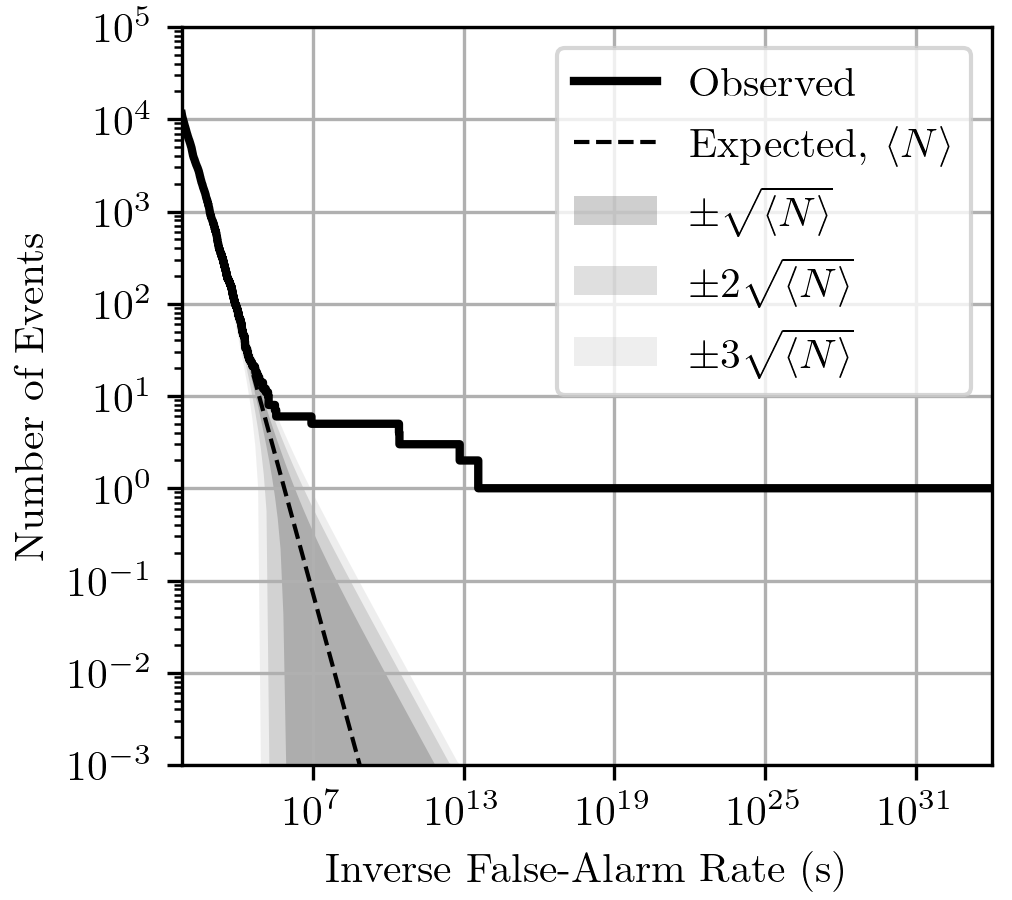}
        \caption{\label{fig:ifar:greg_gpu32}PyTorch GPU float32}
    \end{subfigure}
    \hfill
    \begin{subfigure}{0.49\textwidth}
        \centering
        \includegraphics[width=\textwidth,  keepaspectratio]{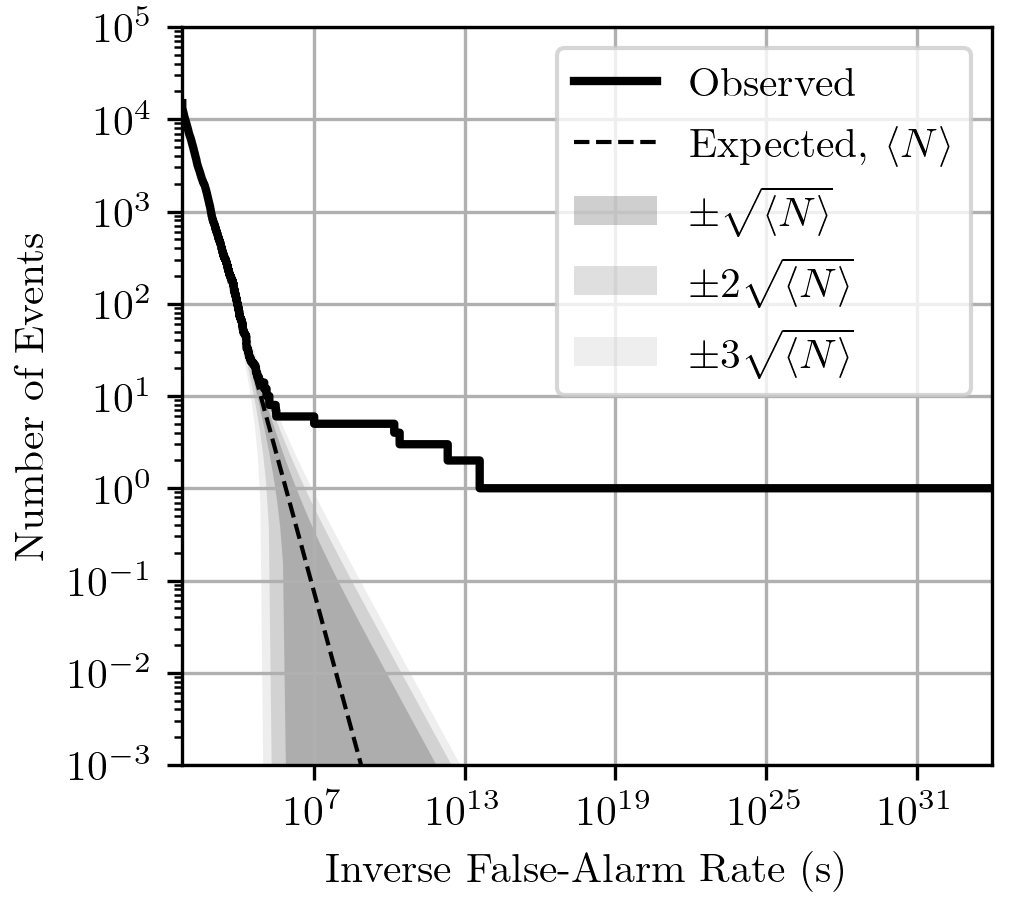}
        \caption{\label{fig:ifar:greg_gpu16}PyTorch GPU float16}
    \end{subfigure}
    \caption{Number of events vs. inverse false-alarm rate for the four analyses. The dashed lines are the expected number of events coming from background noise. The black lines are the number of events actually observed.}
    \label{fig:ifar} 
\end{figure*}

Figure \ref{fig:ifar} shows the inverse false-alarm rate (IFAR) plots for the four different analyses for the duration of the search dataset. The dashed lines represent the expected count distributions from background noise. The observed distributions agree with the expected noise count at low IFAR and clearly branch out at six number of events, indicating the presence of the six known GW events in the dataset. All the plots in Figure \ref{fig:ifar} for all four of the analyses show consistent IFAR distributions. In particular, Figure \ref{fig:ifar:greg_gpu16} demonstrates that using float16 precision in a GW search pipeline is feasible.

\begin{table*}
\caption{\label{table:events} Search results of the GW events previously reported in GWTC-3 within the dataset for GstLAL and three configurations of the PyTorch adaptation. The instruments listed in the ``found inst." column are those which identified the event with a trigger of SNR $\ge$ 4.0. The SNRs in the table are network SNRs.}
\centering
\begin{ruledtabular}
\footnotesize
\begin{tabular}{lcccccccccccccccccccc}
& & \multicolumn{2}{c}{GstLAL}  & \multicolumn{9}{c}{PyTorch adaptation} \\
\multicolumn{3}{c}{ } & & \multicolumn{3}{c}{CPU float32} & 
\multicolumn{3}{c}{GPU float32} & \multicolumn{3}{c}{GPU float16}
\\\cline{3-4}\cline{6-13}
Name & Found Inst.  &  SNR & FAR  &  & SNR & FAR & & SNR & FAR  &  & SNR & FAR  \\
&&&(yrs$^{-1}$) &&&(yrs$^{-1}$) &&&(yrs$^{-1}$) &&& (yrs$^{-1}$)  \\\hline
GW190513\_205428 & H1L1V1 & 12.30 & 4.92 $\times 10^{-6}$ &  & 12.27 & 1.37 $\times 10^{-5}$ &  & 12.27 & 4.49 $\times 10^{-6}$ &  & 12.27 & 1.29 $\times 10^{-5}$ \\
GW190514\_065416 & H1L1 & 8.404 & 3.83 && 8.403 & 3.32 && 8.403 & 2.62 && 8.402 & 3.24 \\
GW190517\_055101 & H1L1 & 10.01 & 0.0011 && 10.01 & 0.00095 && 	10.01 & 0.0010 && 10.01 & 0.0016 \\
GW190519\_153544 & H1L1 & 13.26 & 7.41 $\times 10^{-7}$ && 13.26 & 6.80 $\times 10^{-7}$ && 13.26 & 7.40 $\times 10^{-7}$ && 13.26 & 8.18 $\times 10^{-7}$ \\
GW190521 & H1L1 & 14.57 & 0.0010 && 14.54 & 0.0010 && 14.54 & 0.0011 && 14.54 & 0.0026 \\
GW190521\_074359 & H1L1 & 23.55 & 3.13 $\times 10^{-27}$ && 23.55 & 3.86 $\times 10^{-27}$ && 23.55 & 3.46 $\times 10^{-27}$ && 23.55 & 6.29 $\times 10^{-27}$ \\
\end{tabular}
\end{ruledtabular}
\end{table*}


\subsection{Injection recovery}
In addition to the strain data, we process an identical stretch of strain data with injection frames added to evaluate the relative performance difference of the four pipeline configurations. 
The injection set was generated with source frame component masses between 10 to 100 $M_\odot$, spin z-components between -0.3 to 0.3, and maximum redshift of 3. The injections are distributed uniformly in comoving volume. Injection frames were generated from the injection parameters and inserted into the strain data every 48 seconds, resulting in a total of 15,615 injections through out the duration of the dataset. 

We next focus on comparing the relative number of found injections across the four analyses. An injection is considered ``found" if it has a FAR value below a FAR threshold of one per thirty days. Table \ref{table:found} lists the number of injections found for each of the analyses, categorized by the observing instrument times. The results demonstrate that the injection recovery of the PyTorch adaptation aligns closely with that of GstLAL. Both the CPU float32 version and GPU float32 version have a total number of found injections with a relative difference of less than $1\%$ compared to GstLAL. The GPU float16 configuration, with a total number of detected events 1.13\% lower than GstLAL, shows slightly reduced performance but remains comparable.


\begin{table}
\caption{\label{table:found}Number of found injections for each of the four analyses, categorized by different combinations of observing instruments. The ``Sum" row adds up all the injection numbers from all the different combinations.}
\centering
\begin{ruledtabular}
\begin{tabular}{lrrrrrrrr}


& & GstLAL & & \multicolumn{3}{c}{PyTorch adaptation} \\
& &&&  CPU & GPU & GPU \\
On Inst.  &&&& float32 & float32 & float16 \\
\cline{1-1}
\cline{3-3}
\cline{5-7}
H1L1V1 & & 1996 & & 2017 & 1990 & 1982 \\ 
H1L1   & &  989 & &  989 &  978 &  966 \\
H1V1   & &   99 & &   99 &   99 &   98 \\
L1V1   & &  330 & &  333 &  332 &  327 \\
H1     & &   11 & &   11 &   11 &   10 \\
L1     & &   77 & &   78 &   78 &   79 \\
V1     & &   19 & &   19 &   19 &   19 \\\hline
Sum    & & 3521 & & 3546 & 3507 & 3481 \\
\end{tabular}
\end{ruledtabular}
\end{table}

    


\section{Computational performance}\label{sec:comp}
In this section, we seek to evaluate the computational performance for the different configurations of the PyTorch adaptation, using the GstLAL analyses as a baseline. The part of the pipeline that is benchmarked is the inspiral job, which begins with the ingestion of strain data, and ends with the generation of triggers. We perform a three detector search, with the strain data being three stretches of white gaussian data. The SVD bank used has 1,004 templates, with template duration of 73 seconds. There is one SVD bank for each instrument, since the SVD templates need to be whitened by the PSDs of the instruments. When we process multiple banks in the following tests, we provide the same SVD bank multiple times.

The CPU benchmarking tests are conducted on a computing node equipped with AMD EPYC 7313 Processors, with 32 physical CPU cores in total. We launched 64 inspiral jobs on the computing node in parallel. Each job processes one hour 
of continuous HLV gaussian data, and one SVD bank for each instrument. We assess computational performance by using ``templates in real-time" as a metric, which is defined as
\begin{equation}\label{templates}
    \frac{\text{number of templates}\times \text{data duration (s)}}{\text{number of cores}\times\text{wall time (s)} }
\end{equation}
where wall time is the real-world time it takes for the program to run from start to finish. This ``templates in real-time" metric serves as a measure of how many templates the pipeline can process in real-time on a single CPU core.

The GPU benchmarking tests are performed on two kinds of GPUs, NVIDIA A2 16GB and NVIDIA A100-SXM4-80GB. As explained in Section \ref{sec:pytorch}, the PyTorch adaptation will stack multiple SVD banks during the filtering stage and process the multiple SVD banks in parallel. Therefore, for benchmarking the GPU configurations, we run one inspiral job and vary the number of SVD banks the pipeline processes to evaluate the performance as a function of the number of SVD banks. We use the same SVD bank used for the CPU benchmarking tests. For the GPU ``templates in real-time" metric, we consider the GPU as one processing unit, and set ``number of cores" as one in Equation \ref{templates}.

\begin{figure}
    \centering
    \includegraphics{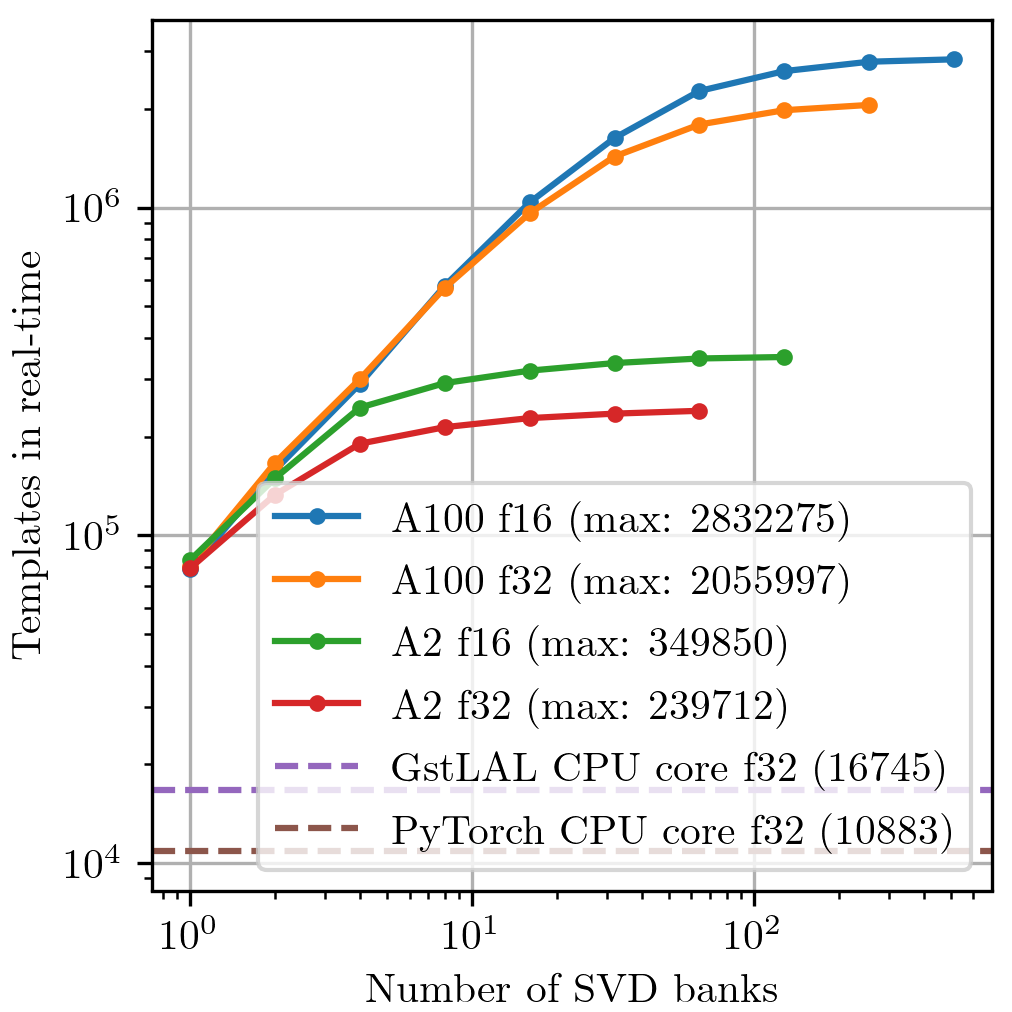}
    \caption{\label{fig:benchmark}Computational performance of the four analyses. For the PyTorch GPU configuration, two different GPUs were used for the benchmarking: NVIDIA A2 16GB (A2) and NVIDIA A100-SXM4-80GB (A100). The purple and brown dashed lines are the templates per CPU core benchmarks for GstLAL and PyTorch CPU configuration, respectively. Float16 and float32 precisions are represented as ``f16" and ``f32", respectively. The numbers in the parenthesis are the maximum templates in real-time values for the GPU benchmarks, and the templates per CPU core in real-time values for the CPU benchmarks.}
\end{figure}

\begin{figure}
    \centering
    \includegraphics{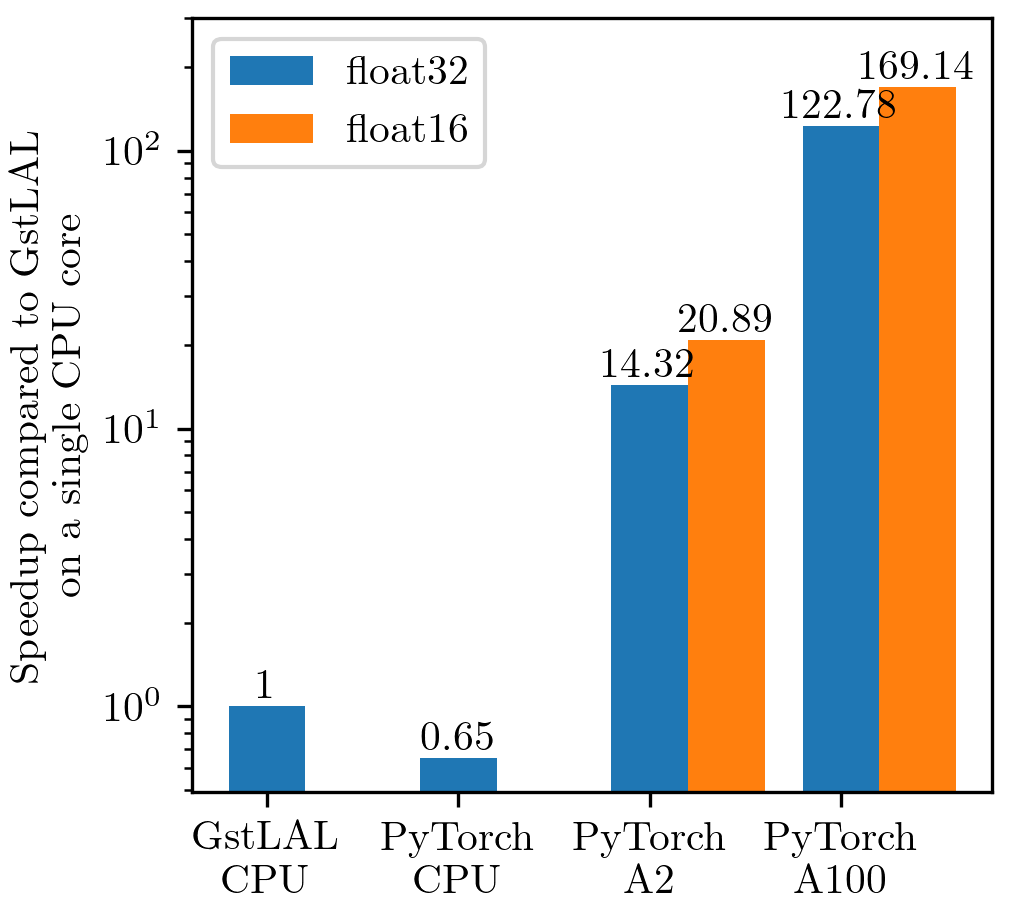}
    \caption{\label{fig:bar}Speedup factor of different device and data type configurations of the PyTorch adaptation compared to GstLAL on a single CPU core. The blue bars represent the configurations with float32 precision, whereas the orange bars represent those with float16 precision.}
\end{figure}

Figure \ref{fig:benchmark} presents the performance of GstLAL and the different configurations of the PyTorch adaptation. The CPU version of PyTorch is slower than that of GstLAL, and its optimization is intended for future work. The performance of the GPU configurations improves as the number of SVD banks processed within one job increases. The performance grows linearly for small number of SVD banks, and slowly plateaus at larger number of SVD banks, when the GPU's resources are saturated. Figure \ref{fig:bar} shows the speedup factor of the PyTorch adaptation compared to GstLAL on a single CPU core. The GPU analyses clearly outperform GstLAL's single CPU core performance, with the highest speedup factor being 169.14x. 


%

\section{Conclusion}\label{sec:conclusion}
In this paper, we present a scalable matched-filter based GW detection pipeline, where the core filtering engine and trigger generator of the GstLAL pipeline have been replaced with PyTorch modules. The flexibility to switch between computation devices allows us to offload computational bottlenecks to GPUs for enhanced performance. Additionally, we introduced a feature in the filtering algorithm that leverages PyTorch's multi-batch and multi-channel options to filter multiple SVD banks in parallel.

To validate the detection performance of the PyTorch adaptation, we filtered GstLAL and three PyTorch configurations: CPU float32, GPU float32, and GPU float16, using an 8.8-day stretch of O3 public data. All four configurations successfully recovered the six known GW events with consistent network SNR and significance. Results from the injection campaign also indicate that injection recoveries were comparable across all analyses.

We benchmarked the PyTorch adaptation and compared its computational performance with GstLAL. Our results show that using the GPU float16 configuration on an A100 GPU, the PyTorch adaptation achieved up to 169 times the speed of GstLAL running on a single CPU core.

This study further demonstrates that float16 precision can yield comparable results to float32 precision, making it effective for detecting known GW events while offering better computational performance. This improvement in efficiency enables us to potentially expand the search parameter space.

In the future, we plan to modernize the remaining portions of the pipeline still using the Gstreamer framework. With the increased performance from GPU configurations, there is potential to further expand the search parameter space. Additionally, leveraging the PyTorch framework will allow us to integrate machine learning techniques, explore alternative detection methods, and enhance the pipeline's overall sensitivity.

\begin{acknowledgments}
This material is based upon work supported by NSF's LIGO Laboratory which is a major facility fully funded by the National Science Foundation.
This research has made use of data, software and/or web tools obtained from the
Gravitational Wave Open Science Center (https://www.gw-openscience.org/ ), a
service of LIGO Laboratory, the LIGO Scientific Collaboration and the Virgo
Collaboration.  We especially made heavy use of the LVK Algorithm
Library~\cite{LAL, lalsuite}. LIGO Laboratory and Advanced LIGO are funded
by the United States National Science Foundation (NSF) as well as the Science
and Technology Facilities Council (STFC) of the United Kingdom, the
Max-Planck-Society (MPS), and the State of Niedersachsen/Germany for support of
the construction of Advanced LIGO and construction and operation of the GEO600
detector. Additional support for Advanced LIGO was provided by the Australian
Research Council.  Virgo is funded, through the European Gravitational
Observatory (EGO), by the French Centre National de Recherche Scientifique
(CNRS), the Italian Istituto Nazionale di Fisica Nucleare (INFN) and the Dutch
Nikhef, with contributions by institutions from Belgium, Germany, Greece,
Hungary, Ireland, Japan, Monaco, Poland, Portugal, Spain.

The authors are grateful for computational resources provided by the 
the Pennsylvania State University's Institute for Computational and Data
Sciences gravitational-wave cluster, and the LIGO Lab cluster at the LIGO Laboratory, and supported by 
the National Science Foundation awards 
OAC-2103662, PHY-2308881, PHY-2011865, OAC-2201445, OAC-2018299, PHY-0757058, PHY-0823459, and PHY-2207728.  
CH Acknowledges generous support from the Eberly College of Science, the 
Department of Physics, the Institute for Gravitation and the Cosmos, and the 
Institute for Computational and Data Sciences.

\end{acknowledgments}

\bibliography{references}

\end{document}